\documentclass[12pt]{article}

\usepackage{theorem}
\usepackage{amsfonts}
\newfont\fiverm{cmr5}

\def\comment#1{{}}	
\theoremstyle{change}
\newtheorem{thm}{Theorem.\nopagebreak}[section]

\newtheorem{prop}[thm]{Proposition.\nopagebreak}

\theorembodyfont{\rmfamily}

\newtheorem{rem}[thm]{Remark.\nopagebreak}

\newtheorem{alg}[thm]{Algorithm:}

\def\bealg#1{\begin{alg}{\bf #1}\index{Algorithm!#1}\nopagebreak}
\def\ealg{\end{alg}}

\def\Box{{\hbox{\raisebox{0.0em}{\rlap{$\sqcap$}}\kern0em%
            \raisebox{-0.0em}{$\sqcup$}}} } 
\newenvironment{proof}{{\it Proof. }}{~~~\hfill$\Box$\vspace{0.5cm}}
\def\bepf{\begin{proof}}
\def\epf{\end{proof}}

\def\fct#1{\mathop{\rm #1}}	
\def\fns#1{{\mbox{\rm \scriptsize#1}}} 		
\def\Lin{\fct{Lin}}
\def\tow{\rightharpoonup}	
\def\re{\fct{Re}}		
\def\im{\fct{Im}}		
\def\eff{{{\fns{eff}}}}
\def\iref{{{\fns{ref}}}}	
\def\tr{\fct{tr}}

\def\D{\displaystyle}				

\def\hbar{{h\hspace{-2mm}^-}}

\def\eps{\varepsilon}
\def\phi{\varphi}

\def\wave{\protect{\footnotesize $\sim$}}

\def\iff{~~~\Leftrightarrow~~~}

\def\downto{\downarrow}

\def\<{\langle} 				
\def\>{\rangle} 				
\def\rangu{\hbox{\raisebox{0.15em}{\rlap{$\sqcap$}}\kern0em%
            \raisebox{-0.19em}{$\sqcup$}} } 

\def\beq{\begin{equation}} 
\def\eeq{\end{equation}} 
\def\lbeq#1{\begin{equation} \label{#1}} 
\def\beqar{\begin{eqnarray}}
\def\eeqar{\end{eqnarray}}
\def\bary{\begin{array}}
\def\eary{\end{array}}
\def\becas{\left\{ \begin{array}{l@{\qquad}l}}
\def\ecas{\end{array} \right.}
\def\benu{\begin{enumerate}}
\def\eenu{\end{enumerate}}
\def\gzit#1{{\rm (\ref{#1})}} 			

\def\Cz{\mathbb{C}}
\def\Hz{\mathbb{H}}

\def\Vz{\mathbb{V}}
\def\Wz{\mathbb{W}}

\def\AC{{L}}

\advance\textheight4.5cm	
\advance\topmargin-2cm

\begin{document}

\vspace*{-2cm}

\begin{center}

{\LARGE \bf Effective Schr\"odinger equations} \\

{\LARGE \bf for nonlocal and/or dissipative systems} \\

\vspace{1cm}

\centerline{\sl {\large \bf Arnold Neumaier}}

\vspace{0.5cm}

\centerline{\sl Institut f\"ur Mathematik, Universit\"at Wien}
\centerline{\sl Strudlhofgasse 4, A-1090 Wien, Austria}
\centerline{\sl email: Arnold.Neumaier@univie.ac.at}
\centerline{\sl WWW: http://www.mat.univie.ac.at/\wave neum/}

\vspace{0.5cm}

\today

\end{center}

\vspace{0.5cm}
{\bf Abstract.} 

The projection formalism for calculating effective Hamiltonians and 
resonances is generalized to the nonlocal and/or nonhermitian case, 
so that it is applicable to the reduction of relativistic systems
(Bethe-Salpeter equations), 
and to dissipative systems modeled by an optical potential. 

It is also shown 
how to recover {\em all} solutions of the time-independent 
Schr\"o\-dinger equation in terms of solutions of the effective 
Schr\"odinger equation in the reduced state space and a 
Schr\"odinger equation in a reference state space. 

For practical calculations, it is important that the resulting 
formulas can be 
used without computing any projection operators. This leads to a 
modified coupled reaction channel/resonating group method framework 
for the calculation of multichannel scattering information.

\vfill

\begin{flushleft}
{\bf Keywords}: Feshbach projection, effective Hamiltonian, 
nonlocal Schr\"odinger equation, Bethe-Salpeter equation,
coupled reaction channels, resonating group method, 
dissipative quantum system, optical potential, form factor, 
doorway operator, time-independent perturbation theory, 
backward error analysis, multichannel scattering\\
\end{flushleft}

\vspace{0.5cm}

{\bf E-print Archive No.}: hep-th/0201085

{\bf 1998\hspace{.4em} PACS Classification}: 
primary: 03.65.-w
secondary: 31.15.Dv  \\

\newpage

\section{Introduction} \label{In}

In many applications, a quantum system of interest is part of a much 
bigger system, and the latter's state influences the system state. 
If the big system is represented by solutions of a Schr\"odinger 
equation in a big state space, it is desirable to find an effective
Schr\"odinger equation in a small state space that describes how
the small system of interest is affected by the embedding into the 
big system. 
Under certain conditions, a Schr\"odinger equation in the big state 
space can indeed be reduced to an effective Schr\"odinger equation 
in the small state space, in such a way that the interesting part of 
any solution of the full Schr\"odinger equation satisfies exactly the
effective Schr\"odinger equation. 

Much effort has gone into solving this reduction problem, and in a 
sense it is well understood \cite{AdhK,FadM,Glo,KukKH,WilT}.
Exact expressions for the effective Hamiltonian can be given. 
In its exact form, the effective Hamiltonian is energy dependent and 
usually acquires a nonhermitian part; its eigenvalues describe the
bound states and resonances of the reduced system.
There are approximation schemes that compute the effective Hamiltonian 
(at least in principle) to arbitrary accuracy.

Less known (but proved here) is that certain solutions of the 
full Schr\"odinger equation can be reconstructed from solutions 
of the effective Schr\"odinger equation, using little more than what 
is already available from the reduction process.

\bigskip
The reduction is usually done for bound state calculations by the 
variational principle discussed in every textbook on quantum mchanics.
For resonance calculations, the reduction may be done within the 
{\sc Feshbach} \cite{Fes1,Fes2} projection formalism (for an
exposition see, e.g., {\sc Kukulin} et al. \cite[Chapter 4]{KukKH}). 
In both cases, the reduced state space is finite-dimensional.
For the calculation of scattering states, the reduction may be done by 
means of coupled reaction channel equations, also called 
the resonating group method; a nice exposition is given in
{\sc Wildermuth \& Tang} \cite{WilT}. In this case,
the reduced state space is a direct sum of finitely many function 
spaces, one for each energetically admissible arrangement of particles 
into clusters, with states parameterized by coordinates or momenta of 
cluster centers only.

The coupled reaction channel equations are numerically easy to handle, 
but the approximations involved in their derivation make an 
assessment of their accuracy difficult. On the other hand, the
projection formalism gives in principle exact results, limited in
accuracy only by the approximations made in the calculation of the
effective Hamiltonian. However, this involves projection operators, 
which are clumsy to use if an orthogonal basis is not easily available.

Moreover, if the full Hamiltonian is already nonhermitian, or if the
Hamiltonian is energy-dependent (such as in relativistic calculations;
cf. \cite{FriM,Ish,IvaLLR,KleFM,KviB,LahA,LuL,MeuGP,RobW}), 
the standard derivation of the 
effective Hamiltonian is no longer valid. 
In the following, we shall remedy both defects.

\bigskip
The paper is organized as follows. We first extend the projection 
formalism such that it applies to the reduction of nonlocal systems
and of dissipative systems 
modeled by an optical potential. Thus we work throughout with complex 
symmetric Hamiltonians, introduced in Section \ref{s.sym}, and 
generalize in Sections \ref{s.red} and \ref{s.pert} the traditional 
Feshbach projection approach to this more general situation. 
We then show in Section \ref{s.schro} that the formalism in fact 
allows to recover {\em all} solutions of the 
time-independent Schr\"odinger equation in terms of solutions of two 
Schr\"odinger equations, one in the reduced state space and the other
in a reference state space. The resulting formulas are closely related 
to those of time-independent perturbation theory.

For practical calculations, the resulting formalism is revised in
Section \ref{s.noproj} so that it can be used without computing any 
projection operators. Section \ref{s.form} shows how the use of
doorway operators gives flexible approximation schemes for the exact
formulas derived in Section \ref{s.noproj}. A backward error analysis 
discussed in Section \ref{s.qual} allows the estimation of the
reliability of approximate solutions of Schr\"odinger equations 
obtained by this or any other method.

Specific choices of the embedding map lead in Section \ref{s.multi} to 
a modified coupled reaction channel/resonating group method framework 
for the calculation of multichannel scattering information, in which
solutions of the full Schr\"odinger equation can be obtained 
from solutions of coupled reaction channel equations for the
effective Hamiltonian. In principle it is capable of arbitrarily 
accurate approximations to the full dynamics, and shares this feature 
with the two Hilbert space method of {\sc Chandler \& Gibson} 
\cite{ChaG,ChaG2}, which partly inspired the present investigations.

The theory is presented in a fully rigorous manner, allowing 
for unbounded operators by using in place of Hilbert spaces a pair of
dual topological vector spaces.
The most useful results are in Sections \ref{s.noproj}, \ref{s.form},
and \ref{s.multi}.

The reader interested in the results but not in mathematical rigor may
omit all references to spaces and topology, may think of all spaces
as finite-dimensional and of operators as matrices, and may skip all 
proofs. In particular, of Section \ref{s.sym} introducing the basic 
terminology, only \gzit{e.hy6a}--\gzit{e.schroe} are essential, 
and most of Sections \ref{s.red}--\ref{s.schro} can be skimmed.

\section{Symmetric operators}\label{s.sym}

$\Lin(\Vz,\Wz)$ denotes the space of continuous linear mappings 
between two topological vector spaces $\Vz$ and $\Wz$, 
$\Vz^*=\Lin(\Vz,\Cz)$ denotes the dual space of continuous, 
complex-valued linear functionals on $\Vz$, and $\Lin\Vz=\Lin(\Vz,\Vz)$ 
denotes the algebra of continuous linear transformations of $\Vz$.

In the following, $\Hz$ is a complex topological vector space with a 
definite, continuous symmetric bilinear form, providing a natural 
embedding of $\Hz$ into the dual space $\Hz^*$.
We refer to $\Hz^*$ as a {\bf state space}, and to the $\psi\in\Hz^*$ 
as {\bf states}. We write the pairing and the bilinear inner product as 
\[
\phi^T\psi=\psi^T\phi~~~\mbox{for }\phi\in\Hz,~\psi\in\Hz^*.
\]
The notation is chosen such that it looks as closely as possible like
standard finite-dimensional linear algebra.

We say that a sequence (or net) $\psi_l\in\Hz^*$ {\bf converges weakly}
to $\psi\in\Hz^*$, and write $\psi_l\tow \psi$, if
\[
\lim_{l\to\infty} \phi^T\psi_l=\phi^T\psi \forall \phi\in\Hz.
\]
We extend the bilinear inner product to arbitrary pairs 
$(\phi,\psi)\in\Hz^*\times \Hz^*$ for which 
\[
\phi^T\psi=\lim_{l\to\infty} \phi_l^T\psi_l
\]
is defined and independent of the weakly converging sequences (or nets)
$\phi_l\tow \phi$, $\psi_l\tow \psi$ of $\phi_l, \psi_l \in\Hz$.
(In the applications, this allows to form the inner product of state 
vectors corresponding to bound states and resonances but not that of 
scattering states.)

Complex conjugation is denoted by a bar, and, with the notation
$\psi^*=\bar\psi^T$, the Hermitian inner product on $\Hz$ is 
\[
\<\phi|\psi\>=\phi^*\psi~~~\mbox{for $\phi,\psi\in\Hz$}.
\]
The associated Euclidean norm is
\[
\|\psi\|=\sqrt{\psi^*\psi},
\]
and $\bar \Hz$ is the closure of $\Hz$ in $\Hz^*$ with respect to the 
Euclidean norm. Thus $\Hz\subseteq \bar\Hz\subseteq \Hz^*$, and 
$\bar \Hz$ is a Hilbert space.

In the applications, $\Hz$ is a space of sufficiently nice functions 
(namely arbitrarily often differentiable, with compact support) 
on some finite- or infinite-dimensional manifold, the 
inner product of two functions is some integral of their pointwise 
product induced by a nonnegative measure on the manifold, 
and $\Hz\subseteq \bar\Hz\subseteq \Hz^*$ is a Gelfand 
triple (or rigged Hilbert space). (For these concepts, see
{\sc Gelfand \& Vilenkin} \cite{GelV}, {\sc Maurin} \cite{Mau}.
For an exposition in physicists' terms see {\sc Kukulin} 
\cite[Appendix A]{KukKH}; cf. also {\sc B\"ohm} \cite{Boh}.)

The {\bf transpose} of a linear operator $A\in\Lin (\Hz,\Hz^*)$ 
is the linear operator $A^T\in\Lin(\Hz,\Hz^*)$ defined by
\[
(A^T\phi)^T\psi=\phi^TA\psi ~~~\mbox{for }\phi,\psi\in\Hz,
\]
$A^*=\bar A^T$ defines the {\bf adjoint} of $A$, and
\[
\re A=\frac{1}{2}(A+A^*),~~~\im A=\frac{1}{2i}(A-A^*)
\] 
define the {\bf real} and {\bf imaginary part} of $A$.
Clearly, $(AB)^T=B^TA^T$ and $(AB)^*=B^*A^*$. 
The operator $A\in \Lin \Hz$ is called {\bf symmetric} if $A^T=A$ on 
$\Hz$, {\bf Hermitian} if $A^*=A$ on $\Hz$, and {\bf positive 
semidefinite} if 
\[
\psi^*A\psi\ge 0~~~\mbox{for all $\psi\in\Hz$}.
\]
In particular, $A$ is Hermitian if and only if $\im A=0$.
We extend symmetric operators $A\in \Lin \Hz$ to $\Lin \Hz^*$ by 
defining $A\psi\in\Hz^*$ for $\psi\in\Hz^*$ by
\[
\phi^TA\psi=(A\phi)^T\psi \forall\phi\in\Hz.
\]

In the following, $\AC \in \Lin \Hz $ is always a symmetric operator 
such that  
\lbeq{e.hy6a}
\im \AC  \mbox{ is positive semidefinite}. 
\eeq
(In particular, this includes the case where $\AC $ is Hermitian
since then $\im \AC =0$.)

Since $\psi^*\AC \psi=\psi^*(\re \AC )\psi+i \psi^*(\im \AC )\psi$ and both
$\psi^*(\re \AC )\psi$ and $\psi^*(\im \AC )\psi$ are real, \gzit{e.hy6a}
is equivalent to 
\lbeq{e.hy6b}
\im \psi^*\AC \psi \ge 0~~~\mbox{for all } \psi\in\Hz.
\eeq
In particular, the spectrum of $\AC $ is in the complex upper half plane. 
If $\AC $ has a spectral resolution then, since 
$|\lambda+i\eps|\ge \eps$ for $\im \lambda\ge 0$, we conclude that 
$(\AC +i\eps)^{-1}$ exists for all $\eps>0$ as a bounded operator on 
$\bar\Hz$ with spectral norm
\[
\|(\AC +i\eps)^{-1}\|\le \eps^{-1}~~~\mbox{for all } \eps>0.
\]
The traditional situation is the one where $\AC =E-H$ with a complex 
energy $E$ satisfying $\im E\ge 0$ and a {\bf Hamiltonian} 
$H=H_s-\frac{i}{2}\Gamma$. Here $H_s,\Gamma\in \Lin \Hz $ are symmetric
and Hermitian, and $\Gamma$ is positive semidefinite. In the most 
important case of a conservative system, $\Gamma=0$ and $H$ is 
Hermitian. However, care is taken that all our results hold in the 
nonhermitian case, corresponding to dissipative systems with an
optical potential that contributes to $\Gamma$. 

The use of $\AC $ helps to avoid a multitude of expressions involving 
$E-H$ or $E-H+i\eps$. Since the (time-independent) 
{\bf Schr\"odinger equation} $H\psi=E\psi$ takes the simple form 
\lbeq{e.schroe}
\AC \psi=0, 
\eeq
this makes the formal manipulations independent of energy and free of 
references to the Hamiltonian $H$, and thus much more readable. 
(Of course, in actual calculations, $E$ and $H$ reappear.) 

More generally, \gzit{e.schroe} also covers nonlocal problems with a 
nonlinear dependence of $\AC $ on $E$, and therefore can be used 
for the Bethe-Salpeter equations arising in bound state and resonance
calculations for relativistic systems \cite{Ish,IvaLLR,LahA,LuL,RobW}.

\section{Reduction of the state space}\label{s.red}

Let $\Hz$ and $\Hz_\eff$ be topological vector spaces with a 
definite, continuous bilinear 
inner product, related by the {\bf embedding map} $P$, an injective, 
closed linear operator from $\Hz_\eff^*$ to $\Hz^*$ satisfying 
$\bar P=P$ and
\[
P\psi_\eff\in\Hz~~~\mbox{for all }\psi_\eff \in \Hz_\eff.
\] 
We want to relate a Schr\"odinger equation $\AC \psi=0$ in the full state 
space $\Hz^*$ to an effective Schr\"odinger equation 
$\AC _\eff\psi_\eff=0$ in the reduced state space $\Hz_\eff^*$.

$P^*=P^T$ maps $\Hz^*$ to $\Hz_\eff^*$ and $\Hz$ to $\Hz_\eff$.
Moreover, $P^*P:\Hz_\eff^*\to\Hz_\eff^*$ is invertible since $P$ is 
closed and injective. The {\bf pseudo inverse} 
\lbeq{re.1}
   P^I:=(P^*P)^{-1}P^*
\eeq
maps $\Hz^*$ to $\Hz_\eff^*$ and possesses the properties
\lbeq{re.1a}
(PP^I)^T=PP^I,~~~P^TPP^I=P^T,~~~P^IP=1.
\eeq
This implies that
\lbeq{re.2}
   Q:=1-PP^I=1-(P^I)^TP^T\in \Lin \Hz^*  
\eeq
satisfies
\lbeq{re.3}
   P^IQ=P^TQ=0,~~~QP=0,~~~Q^2=Q^T=Q.
\eeq
and hence is the orthogonal projection to the orthogonal complement of 
the range of $P$. 

A {\bf $Q$-resolvent} of a symmetric operator 
$\AC \in \Lin \Hz $ is a symmetric operator $G\in \Lin \Hz $ satisfying 
\lbeq{re2.1a}
   GP=0,~~~G\AC Q=Q.
\eeq
Since $GQ=G(1-PP^I)=G$, the symmetry of $G$ implies
\lbeq{re2.1}
   QG=G=GQ,~~~Q\AC G=Q=G\AC Q.
\eeq
Formally, $G=Q(Q\AC Q)^{-1}Q$, but $(Q\AC Q)^{-1}$ is only defined on the 
range $Q\Hz^*$ of $Q$. 

The following discussion generalizes the Feshbach projection formalism 
which is obtained in the special case where $\AC =E-H$ and $H$ is 
Hermitian and $G$ is the ordinary resolvent of $QHQ$ in $Q\Hz^*$
(cf. the development in {\sc Kukulin} et al. \cite[Chapter 4]{KukKH};
in their notation, $G=G_Q(z)$ is called the `orthogonalized resolvent').

\begin{prop}\label{p3.1}~\nopagebreak

(i) If $\AC $ has a $Q$-resolvent $G$, it is uniquely determined,  
the operator
\lbeq{re2.3}
   \AC _\eff:=P^T\AC P-P^T\AC G\AC P=P^T(\AC -\AC G\AC )P=P^T\AC (P-G\AC P)
\eeq
is symmetric, and $\im \AC _\eff$ is positive semidefinite.
  
(ii) For arbitrary $\psi_\eff\in\Hz_\eff^*$, the vector
\lbeq{re2.4}
   \psi=(P-G\AC P)\psi_\eff\in\Hz^*
\eeq
satisfies
\lbeq{re2.4a}
   P^T\AC \psi=\AC _\eff \psi_\eff,~~~Q\AC \psi=0,~~~ \psi_\eff=P^I\psi.
\eeq

\end{prop}

\bepf 
If \gzit{re2.1} holds with $G'$ in place of $G$ then 
\[
G'=G'Q=G'Q\AC G=G'\AC G=G'\AC QG=QG=G, 
\]
giving uniqueness. The vector \gzit{re2.4} satisfies
\[
Q\AC \psi=Q\AC (P-G\AC P)\psi_\eff=(Q-Q\AC G)\AC P\psi_\eff=0,
\]
\[
P^T\AC \psi=P^T\AC (P-G\AC P)\psi_\eff=\AC _\eff\psi_\eff,
\]
and, since $P^IG=P^IQG=0$, 
\[
P^I\psi=P^I(P-G\AC P)\psi_\eff=P^IP\psi_\eff=\psi_\eff.
\]
This proves (ii). Since $\bar P=P$, we have
\[
\bary{lll}
\psi^*\AC \psi&=&\psi^*((P^I)^TP^T+Q)\AC \psi=\psi^*(P^I)^T\AC _\eff \psi_\eff\\
&=&\psi^*(P^I)^*\AC _\eff \psi_\eff=\psi_\eff^*\AC _\eff \psi_\eff.
\eary
\]
Thus, for arbitrary $\psi_\eff\in\Hz_\eff^*$, we have
$\im \psi_\eff^T\AC _\eff \psi_\eff=\im \psi^*\AC \psi \ge 0$.
This proves (i).
\epf

The second term 
\lbeq{e.opt}
\Delta:=P^T\AC G\AC P
\eeq
in the definition of $\AC _\eff$ is called the {\bf optical potential} 
induced by the reduction process. 
(The name is explained in {\sc Taylor} \cite[p.385]{Tay}.)

In the special case where $P^TP=1$ and $\AC =E-H$, $G=G(E)$ and hence the 
optical potential $\Delta(E)=P^T(E-H)G(E)(E-H)P$ is energy-dependent
(and nonlocal) and we have $\AC _\eff=E-H_\eff(E)$ with the 
{\bf effective Hamiltonian}
\[
H_\eff(E)=P^THP+\Delta(E).
\]
Thus the optical potential causes energy shifts in the eigenvalues
of the {\bf projected Hamiltonian} $P^THP$. We also note that the
reduced Schr\"odinger equation is generally a {\em nonlinear} 
eigenvalue problem
\[
H_\eff(E)\psi=E\psi.
\]
Frequently, the energy-dependence is ignored; however, nonlinear 
eigenvalue problems for nonlocal Schr\"odinger equations arising from
Bethe-Salpeter equations for relativistic problems
were actually solved, e.g., \cite{Ish,IvaLLR,LahA,LuL,RobW}.

Note that the resonating 
group method ({\sc Wildermuth \& Tang} \cite{WilT}) works with coupled 
reaction channel equations derived from a projected Hamiltonian and
hence misses the optical potential; an energy-independent term is 
added instead on a phenomenological basis \cite[Chapter 8.2]{WilT}. 
An alternative exact method is derived below.

We return to the general
case.

\begin{prop}
The following identities hold:
\lbeq{re2.8}
   \AC _\eff P^I=P^T(1-\AC G)\AC ,~~~(P^I)^T\AC _\eff =\AC (P-G\AC P),
\eeq
and, if $\AC $ and $\AC _\eff$ are invertible, 
\lbeq{red.12r}
   P^T\AC ^{-1}P=P^TP\AC _\eff^{-1}P^TP.
\eeq
\end{prop}

\bepf
Since $(\AC -\AC G\AC )Q=\AC (Q-G\AC Q)=0$, we have
\[
   \AC _\eff P^I=P^T(\AC -\AC G\AC )PP^I=P^T(\AC -\AC G\AC )(1-Q)=P^T(\AC -\AC G\AC ).
\]
This gives the first equation in \gzit{re2.8}, and the transpose gives
the second equation. For invertible $\AC $, $\AC _\eff$,
we conclude from \gzit{re2.8} and $GP=GQP=0$ that
\[
\AC _\eff P^I\AC ^{-1}P= P^T(1-\AC G)P=P^TP, 
\]
hence
$P^TP\AC _\eff^{-1}P^TP=P^TPP^I\AC ^{-1}P=P^T\AC ^{-1}P$.
\epf

\gzit{red.12r} implies that for invertible $\AC $ and 
$\AC _\eff$, the part of $\AC ^{-1}$ accessible from $\Hz_\eff^*$ can be 
computed from the knowledge of $\AC _\eff$ alone. Similarly, our next
result says that, if $\AC $ has a $Q$-resolvent, all solutions of the
Schr\"odinger equation $\AC \psi=0$ can be computed from 
solutions of the reduced Schr\"odinger equation $\AC _\eff\psi_\eff=0$ 
and a knowledge of the {\bf correction operator} 
\lbeq{re2.8c}
R=G\AC P
\eeq
occurring in \gzit{re2.3} and \gzit{re2.4}.

\begin{thm}\label{t0.1}
Suppose that $\AC $ has a $Q$-resolvent $G$. 

(i) For any $\psi_\eff\in\Hz_\eff^*$ with $\AC _\eff\psi_\eff=0$, the 
vector $\psi\in\Hz^*$ defined by \gzit{re2.4} satisfies $\AC \psi=0$ and
\lbeq{re2.3a}
\psi_\eff=P^I\psi\in\Hz_\eff^*.
\eeq
(ii) For any $\psi\in\Hz^*$ with $\AC \psi=0$, the vector \gzit{re2.3a}
satisfies $\AC _\eff\psi_\eff=0$, and we can reconstruct $\psi$ from
\gzit{re2.4}.

\end{thm}

\bepf 
(i) Multiplication of the second equation of \gzit{re2.8} with 
$\psi_\eff$ gives 
\[
0=(P^I)^T\AC _\eff\psi_\eff=\AC (P-G\AC P)\psi_\eff=\AC \psi, 
\]
and \gzit{re2.3a} follows from Proposition \ref{p3.1}(ii).

(ii) Multiplication of the first equation of \gzit{re2.8} with $\psi$
gives
\[
\AC _\eff\psi_\eff=\AC _\eff P^I\psi=P^T(1-\AC G)\AC \psi=0.
\]
Since $P\psi_\eff=PP^I\psi=(1-Q)\psi=\psi-Q\psi$, \gzit{re2.4}
follows from
\[
\bary{lll}
(1-G\AC )\psi&=&(1-G\AC )\psi-(Q-G\AC Q)\psi=(1-G\AC )(1-Q)\psi\\
&=&(1-G\AC )P\psi_\eff=(P-G\AC P)\psi_\eff
\eary
\]
since
\lbeq{e.gls}
G\AC \psi=0.
\eeq
\epf

Note that, while $\AC _\eff\psi_\eff=0$ looks like a Schr\"odinger
equation, this is generally a {\bf nonlinear} eigenvalue problem.
Indeed, if $\AC =\AC (E)=E-H$ then $\AC _\eff(E)$ is generally a 
nonlinear analytic function of $E$, and the {\bf effective Hamiltonian} 
\[
H_\eff(E):=E-\AC _\eff(E)
\]
has a nonlinear dependence on the energy.

\bigskip
In many cases of interest, the $Q$-resolvent does not exist for the
Hamiltonian $H$ of interest. But frequently the $Q$-resolvents 
$G_\eps$ for the perturbed Hamiltonians $H-i\eps$ corresponding to 
$\AC _\eps=\AC +i\eps$ exist for all $\eps>0$, and are given by
\[
G_\eps=Q(Q\AC Q+i\eps)^{-1}Q.
\]
If the limit
\lbeq{re2.5h}
R:=\lim_{\eps\downto 0} G_\eps \AC _\eps P
\eeq
exists, most of the preceding proof still goes through, 
with $R$ in place of $G\AC P$ and $R^T$ in place of $P^T\AC G$, and
the effective Hamiltonian (usually) acquires a nonhermitian part. 

The only exception is the second half of statement (ii), which must
be modified. (This is most conspicuously seen when $P=0$, where
$\AC _\eff=0$ and the reduced Schr\"odinger equation provides no 
information at all.) Inspection of the proof shows that \gzit{e.gls}
fails. Thus, if $\psi\in\Hz^*$ satisfies $\AC \psi=0$, the expression
\lbeq{re2.5i}
\psi^\perp:=\psi-(P-R)\psi_\eff
\eeq
need not vanish (as predicted by \gzit{re2.4} under the stronger
assumptions), but only the much weaker equation $P^T\psi^\perp=0$
follows. Thus the reduction does no
longer allow one to recover {\em all} solutions of $\AC \psi=0$.
By Theorem \ref{t0.1}(i), $\psi^\perp$ is also a solution of 
the Schr\"odinger equation in $\Hz^*$, and since $P^T\psi=0$ implies 
$\psi_\eff=0$ and hence $\psi^\perp=\psi$ we have
\[
\{\psi^\perp \mid \AC \psi=0\}=\{\psi\mid \AC \psi=0,~P^T\psi=0\}.
\]
We discuss later (Theorem \ref{t4.4}) how to access this missing part.

\section{Perturbation theory}\label{s.pert}

The computation of the $Q$-resolvent is traditionally done using
perturbation theory. It is assumed that the inhomogeneous 
Schr\"odinger equation for a related reference problem is explicitly 
solvable, and the problem of interest is considered as a perturbation
of the reference Schr\"odinger equation.

We therefore assume that we have a symmetric operator 
$\AC _\iref\in \Lin \Hz $ satisfying
\lbeq{re2.14a}
   \AC _\iref Q=Q\AC _\iref;
\eeq
this commutation relation can always be achieved by taking an arbitrary symmetric approximation $\AC _0$ to $\AC $ and putting
\lbeq{re2.22}
   \AC _\iref=\AC _0-(1-Q)\AC _0Q-Q\AC _0(1-Q).
\eeq 
Let $G$ be a $Q$-resolvent of $\AC $ and put
\lbeq{re2.14}
   V=\AC _\iref-\AC ,~~~T=V+VGV,~~~\Omega:=1+GV.
\eeq
$V$ is called the {\bf interaction}, $T$ the ($Q$-version of the)
{\bf transition operator} or {\bf T-matrix}, and $\Omega$ the 
($Q$-version of the) {\bf M\"oller operator}. This
terminology is justified by the close formal relations of the 
properties of $T$ and $\Omega$, derived below, with those of the 
T-matrix and the M\"oller operators of standard scattering theory.
Indeed, for $P=0$, $Q=1$ (which is uninteresting from the point of 
view of effective Schr\"odinger equations), the results reduce to
those of standard perturbation theory.

\begin{prop}\label{p0.3}
If \gzit{re2.14a} holds then 
\lbeq{re2.15a}
   \Omega P=(1-G\AC )P,~~~\Omega^TP=P,
\eeq
\lbeq{re2.15b}
   T=V\Omega=\Omega^TV,
\eeq
\lbeq{re2.15}
   P^T\AC \Omega P=\AC _\eff=P^T(\AC _\iref-T)P.
\eeq
\end{prop}

\bepf
Since 
\[
G\AC _\iref P=GQ\AC _\iref P=G\AC _\iref QP=0
\]
we have
\[
   G\AC P=G\AC _\iref P-GVP=-GVP,
\]
\[
   P^T\AC G=(G\AC P)^T=-(GVP)^T=-P^TVG,
\]
hence 
\[
\Omega P=P+GVP=(1-G\AC )P.
\]
Together with $\Omega^TP=(1+VG)P=(1+VGQ)P=P$, this gives \gzit{re2.15a}.
\gzit{re2.15b} follows directly from \gzit{re2.14}.
Since
\[
\bary{lll}
P^T\AC \Omega P&=&P^T\AC (1-G\AC )P=\AC _\eff\\
&=&P^T\AC P-P^T\AC G\AC P=P^T\AC P+P^T\AC GVP\\
&=&P^T(\AC _\iref-V)P-P^TVGVP=P^T(\AC _\iref-T)P,
\eary
\]
\gzit{re2.15} follows. 
\epf

\begin{prop}\label{p0.3a}
If $\Omega$ is invertible then 
\lbeq{re2.18}
   G_\iref=\Omega^{-1}G
\eeq
is a $Q$-resolvent of $\AC _\iref$,
and 
\lbeq{re2.21}
   \Omega=(1-W)^{-1}, ~~~\mbox{where } W=G_\iref V.
\eeq
\end{prop}

\bepf
\[
\Omega(1-W)=\Omega-\Omega G_\iref V=\Omega-GV=1, 
\]
gives \gzit{re2.21}. Since $G_\iref P=\Omega^{-1}GP=0$ and
\[
G_\iref \AC _\iref Q=G_\iref \AC Q+G_\iref VQ=\Omega^{-1}G\AC Q +WQ
=(1-W)Q+WQ=Q,
\]
$G_\iref$ is a $Q$-resolvent of $\AC _\iref$.
\epf

\begin{thm}\label{t3.4}
Suppose that $G_\iref$ is a $Q$-resolvent of a symmetric operator 
$\AC _\iref$ satisfying \gzit{re2.14a}, and suppose that $\Omega$ 
with \gzit{re2.21} exists. Then $G=\Omega G_\iref$ is a $Q$-resolvent 
of $\AC =\AC _\iref-V$, and with
\lbeq{re2.14b}
   T=V+VGV,
\eeq
\gzit{re2.14}--\gzit{re2.18} hold. Moreover, we have 
\lbeq{red.10}
   \Omega=1+W\Omega=1+\Omega W,
\eeq
\lbeq{red.11}
   T=V+TW=V+W^TT,
\eeq
\lbeq{red.11a}
   G=G_\iref+W G=G_\iref+GW^T.
\eeq
\end{thm}

\bepf
\gzit{red.10} follows directly from $\Omega(1-W)=(1-W)\Omega=1$
and implies $\Omega=1+\Omega G_\iref V=1+GV$, hence \gzit{re2.14}.
Since $GP=\Omega G_\iref P=0$ and
\[
G\AC Q=G\AC _\iref Q-GVQ=\Omega G_\iref \AC _\iref Q-GVQ =\Omega Q-GVQ=Q,
\]
$G$ is a $Q$-resolvent of $\AC $. Thus Proposition \ref{p0.3} applies,
and gives \gzit{re2.15a}--\gzit{re2.15}. \gzit{re2.18} is obvious. 
The first equality in \gzit{red.11} follows from \gzit{re2.15b} and 
\gzit{red.10}, and the second follows by transposing the first 
equation. The first equation in \gzit{red.11a} follows from
\[
WG=G_\iref VG=G_\iref QVQG=G_\iref(Q\AC _\iref Q-Q\AC Q)G=G-G_\iref,
\]
and the second follows again by transposing the first equation.
\epf

Inserting the definition \gzit{re2.21} of $W$ into \gzit{red.11a}
gives the (Q-version of the) {\bf Dyson equation} 
\lbeq{red.11b}
   G=G_\iref+G_\iref V G=G_\iref+GV^T G_\iref.
\eeq

If the spectral norm of $W=G_\iref V$ is smaller than one, 
\gzit{red.10}--\gzit{red.11a} can be used to calculate 
iteratively the M\"oller operator $\Omega$, the transition operator $T$ 
and the $Q$-resolvent $G$. To lowest order, we get 
\[
\Omega\approx 1+G_\iref V,~~~T\approx V,~~~
G\approx G_\iref+G_\iref VG_\iref.
\]
When $\Hz_\eff=\{0\}$, this is the {\bf Born approximation}, and 
when $\Hz_\eff$ is finite-dimensional, this is a
version of the {\bf distorted wave Born approximation} 
(see, e.g., {\sc Newton} \cite[Section 9.1]{New}). 
In the Born approximation, we simply get
\[
\AC _\eff \approx P^T(\AC _\iref -V)P = P^T\AC P;
\]
in second order, 
\lbeq{red.opt}
\AC _\eff \approx P^T\AC P -P^TVG_\iref VP,
\eeq
giving the approximation $\Delta\approx P^TVG_\iref VP$ for the 
optical potential \gzit{e.opt}. 
Further iteration gives the {\bf Born series}
\[
T=V+VG_\iref V+VG_\iref VG_\iref V+\dots,
\]
giving the exact optical potential
\[
\Delta= P^TVG_\iref VP+P^TVG_\iref VG_\iref VP+\dots .
\]

Thus we have recovered a generalized version of traditional 
perturbation theory. For $\Hz_\eff=\{0\}$, $P=0$, $Q=1$, 
the equations obtained above 
reduce to those of standard perturbative scattering theory; 
the missing part -- that one gets the scattering solutions of 
the Schr\"odinger equations -- follows in the next section. 
Of course, from the point of view of 
effective Schr\"odinger equations, the case $P=0$ is completely 
uninteresting; however, it is instructive in that it shows that the
methods used to solve scattering problems apply with small 
modifications to the problem of finding effective Hamiltonians.

Similarly, if $\Hz_\eff$ is the eigenspace of $H_\iref$ corresponding 
to a bound state of the reference system
and $P$ the orthogonal projector to this space,
we get the situation leading (in the Born approximation) to
Fermi's Golden Rule; cf. {\sc Kukulin} et al. \cite[Section 4.4]{KukKH}.
For a nondegenerate bound state, $\Hz_\eff$ is one-dimensional,
and again the point of view of effective Schr\"odinger equations
is empty. For degenerate bound states, however, we recover a
nontrivial low-dimensional eigenvalue problem as effective 
Schr\"odinger equation.

\section{Solving the Schr\"odinger equation}\label{s.schro}

We are now ready to express all solutions of the Schr\"odinger 
equation in terms of the solutions of a reduced Schr\"odinger equation 
and special solutions of a reference Schr\"odinger equation.

\begin{prop}\label{p5.1}
If 
\lbeq{Rs.1}
   \psi_\eff=P^I\psi,~~~\psi_\iref=Q(1-W)\psi
\eeq
then 
\lbeq{Rs.2}
   P^T\psi_\iref=0,~~~\psi=\Omega(P\psi_\eff+Q\psi_\iref).
\eeq
If \gzit{Rs.2} holds then 
\lbeq{Rs.3}
   \AC _\eff\psi_\eff=P^T\Omega^T\AC \psi,
\eeq
\lbeq{Rs.4}
   \AC _\iref\psi_\iref=Q\AC \psi,
\eeq
\lbeq{Rs.5}
   \AC \psi=(P^I)^T\AC _\eff\psi_\eff+\AC _\iref\psi_\iref-
   (1-Q)\Omega^T\AC _\iref\psi_\iref.
\eeq
\end{prop}

\bepf
Since 
\[
   QW=QG_\iref V=G_\iref V=W,
\]
\gzit{Rs.1} implies 
\[
  \bary{ll}
   Q\psi_\iref &=Q(1-W)\psi=(Q-W)\psi\\
   &=(1-PP^I-W)\psi=(1-W)\psi-P\psi_\eff,
  \eary
\]
hence
\[
   \Omega(P\psi_\eff+Q\psi_\iref)=\Omega(1-W)\psi=\psi,
\]
\[
   P^T\psi_\iref=P^TQ(1-W)\psi=0.
\]
This gives \gzit{Rs.2}.
Now suppose that \gzit{Rs.2} holds.
By \gzit{re2.15a},
\lbeq{Rs.6}
   P^T\Omega^T\AC =(\Omega P)^T\AC =P^T(1-\AC G)\AC =P^T\AC (1-G\AC ),
\eeq
and 
\lbeq{Rs.7}
   P^T\Omega^T\AC Q=P^T\AC (Q-G\AC Q)=0.
\eeq
Since $(1-QW)\Omega=(1-W)\Omega=1$, this implies
\lbeq{Rs.8}
   P^T\Omega^T\AC \Omega=P^T\Omega^T\AC (1-QW)\Omega=P^T\Omega^T\AC .
\eeq
Using \gzit{Rs.2}, \gzit{Rs.8}, \gzit{Rs.7}, and \gzit{Rs.6}, we find
\[
  \bary{ll}
   P^T\Omega^T\AC \psi &=P^T\Omega^T\AC \Omega(P\psi_\eff+Q\psi_\iref)
   =P^T\Omega^T\AC (P\psi_\eff+Q\psi_\iref)\\
   &=P^T\Omega^T\AC P\psi_\eff=P^T\AC (1-G\AC )P\psi_\eff=\AC _\eff\psi_\eff,
  \eary
\] 
giving \gzit{Rs.3}.
Since $Q\AC _\iref W=Q\AC _\iref G_\iref V=QV$, we have
\[
   Q\AC \Omega=(Q\AC _\iref-QV)\Omega=Q\AC _\iref(1-W)\Omega=Q\AC _\iref=\AC _\iref Q,
\]
hence
\[
  \bary{ll}
   Q\AC \psi &=Q\AC \Omega(P\psi_\eff+Q\psi_\iref)=\AC _\iref 
   Q(P\psi_\eff+Q\psi_\iref)\\
   &=\AC _\iref Q\psi_\iref=\AC _\iref\psi_\iref,
  \eary
\]
by \gzit{Rs.1}, giving \gzit{Rs.4}.
Finally, $X:=1-(1-Q)\Omega^T$ satisfies
\[
   XP=P-(1-Q)\Omega^TP=P-(1-Q)P=QP=0
\]
by \gzit{re2.15a}, hence
\[
\bary{lll}
   XQ\AC =X(1-PP^I)\AC =X\AC =(1-(1-Q)\Omega^T)\AC =\AC -(P^I)^TP^T\Omega^T\AC ,
\eary
\]
so that
\[
\AC \psi=((P^I)^TP^T\Omega^T+XQ)\AC \psi
=(P^I)^T\AC _\eff\psi_\eff+X\AC _\iref\psi_\iref
\]
by \gzit{Rs.3} and \gzit{Rs.4}.
\epf

Since the formulas in Proposition \ref{p5.1} and $\AC _\eff=P^T\AC \Omega P$ 
do not involve $G$, we can take limits and obtain:

\begin{thm}\label{t4.4}
Let $G_{\iref,\eps}$ be a $Q$-resolvent of $\AC _{\iref,\eps}$ with
$\D\lim_{\eps\downto 0} \AC _{\iref,\eps}=\AC _\iref$. Suppose that 
\[
W:=\lim_{\eps\downto 0} G_{\iref,\eps}V ~~\mbox{and}~~ \Omega=(1-W)^{-1}
\]
exist. If
\[
   \psi_\eff=P^I\psi,~~~\psi_\iref=Q(1-W)\psi
\]
then 
\[
   P^T\psi_\iref=0,~~~\psi=\Omega(P\psi_\eff+Q\psi_\iref).
\]
Moreover, with $\AC _\eff:=P^T\AC \Omega P$, we have
\[
 \AC \psi=0 \iff \AC _\eff\psi_\eff=0,~~~\AC _\iref\psi_\iref=0.
\]
\end{thm}

\section{State space reduction without projections}\label{s.noproj}

Projection operators and the associated $Q$-resolvents are clumsy to 
use if an orthogonal basis is not easily available. 
We therefore revise the above formalism 
so that it can be used without computing any projection operators
or $Q$-resolvents.
Additional flexibility is gained by using in place of the reference 
operator $\AC _\iref\in \Lin \Hz $ an appropriate operator 
$P_0:\Hz_0^*\to\Hz^*$ from the dual of a reference space 
$\Hz_0$.
For exact solutions, this reference space must be at least as big as 
$\Hz$; however, in Section \ref{s.form}, we choose $\Hz_0$ to be a 
smaller space in which numerical calculations are tractable, and obtain
practical approximation schemes for the correction operator and the
effective Hamiltonian.
(This is related to the two Hilbert space method of 
{\sc Chandler \& Gibson} \cite{ChaG,ChaG2}, and indeed was inspired by 
their work.)

As before, $\AC \in \Lin \Hz $ is assumed to be symmetric, and 
$P:\Hz_\eff^*\to\Hz^*$ is assumed to be a closed injective linear 
mapping satisfying $\bar P=P$ and
\[
P\psi_\eff\in\Hz~~~\mbox{for all }\psi_\eff \in \Hz_\eff.
\] 
Let $P_0:\Hz_0^*\to\Hz^*$ be a closed, surjective linear mapping
satisfying 
\[
P_0\psi_0\in\Hz ~~~\mbox{for all }\psi_0 \in \Hz_0.
\] 
Then $P_0^T$ maps $\Hz^*$ to $\Hz_0^*$ and $\Hz$ to $\Hz_0$.
Moreover, $P_0P_0^T:\Hz^*\to\Hz^*$ is invertible since $P_0$ is 
closed and surjective. 

The pseudo inverse $P^I$ of the injective $P$ satisfies as before
\[
P^I=(P^TP)^{-1}P^T,~~~P^IP=1,~P^TPP^I=P^T
\]
but the pseudo inverse $P_0^I$ of the surjective $P_0$ satisfies
\[
P_0^I=P_0^T(P_0P_0^T)^{-1},~~~P_0P_0^I=1,~P_0^IP_0P_0^T=P_0^T.
\]

\begin{prop}
If there are linear mappings $R:\Hz_\eff^*\to\Hz^*$ and 
$\AC ':\Hz_\eff^*\to\Hz_\eff^*$ such that
\[
R\psi_\eff\in\Hz ~~\mbox{for all }\psi_\eff \in \Hz_\eff;~~~~~
\AC '\psi_0\in\Hz_\eff ~~\mbox{for all }\psi_0 \in \Hz_\eff
\] 
and 
\lbeq{pr.37}
P_0^T\AC R+P_0^TP\AC '=P_0^T\AC P,~~~P^TR=0,
\eeq
then
\lbeq{pr.38}
   \AC _\eff:=P^T\AC P-R^T\AC R\in \Lin \Hz 
\eeq
is a symmetric operator satisfying
\lbeq{pr.39}
   \AC _\eff=P^T\AC (P-R),~~~(P^I)^T\AC _\eff=\AC (P-R).
\eeq
Moreover, if $\AC $ and $\AC _\eff$ are invertible then
\lbeq{red.12s}
   P^T\AC ^{-1}P=P^TP\AC _\eff^{-1}P^TP.
\eeq

\end{prop}

\bepf
Symmetry is obvious. 
Multiplication of \gzit{pr.37} with $Q(P^I)^T$
gives 
\[
Q\AC P=Q\AC R+QP\AC '=Q\AC R. 
\]
Now
\[
\bary{lll}
R^T\AC R&=&R^T\AC (Q+(P^I)^TP^T)R=R^T\AC QR\\
&=&(Q\AC R)^TR=(Q\AC P)^TR=P^T\AC QR=P^T\AC R
\eary
\]
since $QR=(1-(P^I)^TP^T)R=R$, hence
\[
\AC _\eff=P^T\AC P-R^T\AC R=P^T\AC P-P^T\AC R=P^T\AC (P-R).
\]
This is the first half of \gzit{pr.39}. By multiplication with 
$(P^I)^T$, we find
\[
\bary{lll}
(P^I)^T\AC _\eff&=&(P^I)^TP^T\AC (P-R)=(1-Q)\AC (P-R)\\
&=&\AC (P-R)-Q\AC P+Q\AC R=\AC (P-R)
\eary
\]
giving the second half of \gzit{pr.39}.
If $\AC $ and $\AC _\eff$ are invertible then \gzit{pr.39} implies
\[
P^T\AC ^{-1}(P^I)^T\AC _\eff=P^T(P-R)=P^TP,
\]
hence
\[
   P^TP\AC _\eff^{-1}P^TP=P^T\AC ^{-1}(P^I)^TP^TP=P^T\AC ^{-1}P,
\]
giving \gzit{red.12s}.
\epf

It is not difficult to see that $R$ and $\AC _\eff$ from 
Section \ref{s.red} are an instance of this construction, with
$\Hz_0=\Hz$ and $P_0=1$.

\begin{prop}~

(i) For arbitrary $\psi\in \Hz^*$, the vectors 
\lbeq{pr.40}
\psi_\eff=P^I\psi\in \Hz_\eff^*,~~~
\psi_0=P_0^I(\psi-(P-R)\psi_\eff)\in \Hz_0^*
\eeq
satisfy
\lbeq{pr.41}
   P^TP_0\psi_0=0,~~~\psi=(P-R)\psi_\eff+P_0\psi_0.
\eeq
(ii) If \gzit{pr.41} holds for some $\psi_\eff\in \Hz_\eff^*$,
$\psi_0\in \Hz_0^*$ then 
\lbeq{pr.42}
   \AC _\eff\psi_\eff=(P-R)^T\AC \psi,
\eeq
\lbeq{pr.43}
   P_0^T\AC P_0\psi_0=(P_0-(P-R)P^IP_0)^T\AC \psi,
\eeq
\lbeq{pr.44}
   \AC \psi=(P^I)^T\AC _\eff\psi_\eff+(P_0^I)^T(P_0^T\AC P_0\psi_0).
\eeq
\end{prop}

\bepf
(i) follows from
\[
P_0\psi_0=P_0P_0^I(\psi-(P-R)\psi_\eff)=\psi-(P-R)\psi_\eff,
\]
\[
\bary{lll}
P^TP_0\psi_0&=&P^T\psi-P^T(P-R)\psi_\eff=P^T\psi-P^TP\psi_\eff\\
&=&P^T\psi-P^TPP^I\psi=0.
\eary
\]
(ii) Since by \gzit{pr.37}
\[
\bary{lll}
R^T\AC \psi&=& R^T\AC P_0\psi_0=(P_0^T\AC R)^T\psi_0
=(P_0^T\AC P-P_0^TP\AC ')^T\psi_0\\
&=&P^T\AC P_0\psi_0-\AC '^TP^TP_0\psi_0=P^T\AC P_0\psi_0,
\eary
\]
\gzit{pr.42} follows from \gzit{pr.39} and \gzit{pr.41}:
\[
\bary{lll}
\AC _\eff\psi_\eff&=&P^T\AC (P-R)\psi_\eff=P^T\AC (\psi-P_0\psi_0)\\
&=&P^T\AC \psi-P^T\AC P_0\psi_0=P^T\AC \psi-R^T\AC \psi=(P-R)^T\AC \psi.
\eary
\]
\gzit{pr.43} holds since by \gzit{pr.41}, \gzit{pr.39} and 
\gzit{pr.42},
\[
\bary{lll}
P_0^T\AC P_0\psi_0&=&P_0^T\AC \psi-P_0^T\AC (P-R)\psi_\eff
=P_0^T\AC \psi-P_0^T(P^I)^T\AC _\eff\psi_\eff\\
&=&P_0^T\AC \psi-P_0^T(P^I)^T(P-R)^T\AC \psi\\
&=&(P_0-(P-R)P^IP_0)^T\AC \psi.
\eary
\]
\gzit{pr.44} holds since by \gzit{pr.40} and \gzit{pr.39},
\[
\bary{lll}
(P_0^I)^TP_0^T\AC P_0\psi_0&=&\AC P_0\psi_0=\AC P_0P_0^I(\psi-(P-R)\psi_\eff)\\
&=&\AC (\psi-(P-R)\psi_\eff)=\AC \psi-(P^I)^T\AC _\eff\psi_\eff.
\eary
\]
\epf

We now have the following projector-free and $Q$-resolvent-free 
version of Theorem \ref{t4.4},
constructing all solutions of the Schr\"odinger equation $\AC \psi=0$ in 
terms of two simpler Schr\"odinger equations.

\begin{thm} ~

(i) If $\im \AC $ is positive semidefinite then $\im \AC _\eff$ is 
positive semidefinite. If also $\bar P_0=P_0$ then also 
$\im P_0^T\AC P_0$ is positive semidefinite.

(ii) For arbitrary $\psi\in \Hz^*$, $\psi_\eff\in \Hz_\eff^*$,
$\psi_0\in \Hz_0^*$ satisfying \gzit{pr.40} or \gzit{pr.41},
\[
 \AC \psi=0\iff \AC _\eff\psi_\eff=0,~P_0^T\AC P_0\psi_0=0.
\]
In particular, to find all solutions of $\AC \psi=0$, it suffices to solve the two problems
\[
\AC _\eff\psi_\eff=0,
\]
\[
P_0^T\AC P_0\psi_0=0,~~~ P^TP_0\psi_0=0.
\]
\end{thm}

\bepf
(i) For arbitrary $\psi_\eff\in \Hz_\eff^*$, we define $\psi$ by 
\gzit{pr.41} with $\psi_0=0$. Then 
$P^T\psi=P^T(P-R)\psi_\eff=P^TP\psi_\eff$, hence $\psi_\eff=P^I\psi$.
Now \gzit{pr.44} gives 
\[
\AC \psi=(P^I)^T\AC _\eff\psi_\eff,
\]
and since $(P^I)^*=(P^I)^T$, we get
\[
\psi^*\AC \psi=\psi^*(P^I)^T\AC _\eff\psi_\eff
=(P^I\psi)^*\AC _\eff\psi_\eff=\psi_\eff^*\AC _\eff\psi_\eff.
\]
Hence $\psi_\eff^*(\im \AC _\eff)\psi_\eff=\psi^*(\im \AC )\psi\ge 0$, and
$\im \AC _\eff$ is positive semidefinite. If also $\bar P_0=P_0$ then 
$P_0^*=P_0^T$ and $\im P_0^T\AC P_0$ is positive semidefinite since
\[
\psi_0^*(\im P_0^T\AC P_0)\psi_0=(P_0\psi_0)^*(\im \AC )(P_0\psi_0)\ge 0.
\]
(ii) The forward implication follows directly from \gzit{pr.42} and 
\gzit{pr.43}, the reverse implication from \gzit{pr.44}.
\epf

\begin{rem}
If $\psi_0=0$ then $\phi^T\psi=\phi_\eff^T(P^TP+R^TR)\psi_\eff$,
so that 
\[
G_\eff=P^TP+R^TR
\]
is the {\bf effective metric} induced on $\Hz_\eff$. 
Note that it is generally not the original metric in  $\Hz_\eff$, 
not even when $P$ is an orthogonal projection.

\end{rem}

\section{Form factors}\label{s.form}

In this section we show how to obtain efficiently approximations 
to the correction operator $R$.
This leads in Section \ref{s.multi} to a modified 
coupled reaction channel/resonating group method framework 
for the calculation of multichannel scattering information.

We emphasize that the formulas derived in this section no longer 
involve a pseudo inverse. In particular, they can be used even when 
$P$ is not injective and $P_0$ is not surjective. 
(Hovever, since the assumptions under which the formulas are derived 
are then violated, they lose some information and hence give only 
approximate effective Hamiltonians.)

\begin{thm}\label{t.form}
Let 
\lbeq{fo.1}
   \AC _0:=P_0^T\AC P_0,~~~P_1:=P^TP_0,~~~U:=P_0^T\AC P-P_0^TP\widetilde \AC 
\eeq
with a symmetric $\widetilde \AC \in \Lin \Hz_\eff$ (and hence in 
$\Lin \Hz_\eff^*$), and write
\[
   \widehat \AC _\eps:=\left(\bary{cc}
                  \AC _\eps & P_1^T\\
                  P_1 &-i\eps
                 \eary\right)\in \Lin (\Hz_0\oplus\Hz_\eff)~~~
\mbox{with }\AC _\eps=\AC _0+i\eps.
\]
(i) If the strong limit
\lbeq{fo.2}
\lim_{\eps\downto 0}\widehat \AC _\eps^{-1}{U\choose 0}={F_0\choose F_1}
\eeq
exists then \gzit{pr.37} is solved by
\lbeq{fo.3}
   R=P_0F_0,~~~\AC '=\widetilde \AC +F_1.
\eeq
(ii) Relation \gzit{fo.2} holds with
\lbeq{fo.3e}
F_0=\lim_{\eps\downto 0} F_{0\eps},~~~
F_1=\lim_{\eps\downto 0} F_{1\eps},
\eeq
where 
\lbeq{fo.3f}
F_{1\eps}=(P_1\AC _\eps^{-1}P_1^T+i\eps)^{-1}P_1\AC _\eps^{-1}U,~~~
F_{0\eps}=\AC _\eps^{-1}(U-P_1^TF_{1\eps}),
\eeq
if these limits exist.
\end{thm}

\bepf
(i) follows from 
\[
  \bary{ll}
   \D{P_0^T\AC R+P_0^TP\AC '\choose P^TR}
   &=\D{P_0^T\AC P_0F_0+P_0^TPF_1+P_0^TP\widetilde \AC \choose P^TP_0F_0}\\
   &=\D\widehat \AC _0{F_0\choose F_1}+{P_0^TP\widetilde \AC \choose 0}\\
   &=\D\lim_{\eps\downarrow 0}\widehat \AC _0\widehat \AC _\eps^{-1}
   {U\choose 0}+{P_0^TP\widetilde \AC \choose 0}\\
   &=\D{U\choose 0}+{P_0^TP\widetilde \AC \choose 0}={P_0^T\AC P\choose 0}
   \eary
\]
by looking at the upper and the lower part separately.
(ii) follows from the equation
\[
\widehat \AC _\eps{F_{0\eps}\choose F_{1\eps}}={U\choose 0},
\]
which is easily verified by substitution.
\epf

In principle, $\widetilde \AC $ in \gzit{fo.1} may be chosen arbitrarily.
However, for numerical calculations it may be advisable to choose 
$\widetilde \AC $ in such a way that $U$ (which replaces
the interaction $V$ in the projection approach) becomes small in some 
sense. This has the beneficial consequence that then the numerical
approximation errors have a much smaller effect on the calculated 
solution. 

\bigskip
In practice it is impossible to compute the exact correction operator
and hence the exact effective Hamiltonian, since these tend to be 
exceedingly complicated. One therefore exploits physical intuition 
to select a space $\Hz_0$ of manageable complexity whose dual contains 
the {\bf doorway states} believed to mediate the interaction of the 
reduced system and the unmodelled environment. $\Hz_0^*$ is embedded
into $\Hz^*$ by means of a {\bf doorway operator} $P_0$ that is now 
no longer surjective. Fortunately, the formulas 
\gzit{pr.37} and \gzit{pr.38} defining $R$ and $\AC _\eff$ do not depend
on pseudo inverses, and hence make also sense in this case. 
\gzit{pr.37} and \gzit{pr.38} now only yield approximate solutions for
the correction operator and an approximate effective Hamiltonian.
However, these approximations become better and better as the range of 
$P_0$ covers a bigger and bigger part of $\Hz$. 

The situation is fully analogous to numerical discretization schemes 
that are necessary to solve all but the simplest partial differential 
equations; the only difference is that in the present context it 
frequently makes sense to consider approximations in manageable 
function spaces, so that one does not discretize completely.

A proper choice of the doorway operator $P_0$ makes the computations 
more tractable. At the same time, it limits the formal complexity of 
the optical potential
\lbeq{fo.4}
   \Delta=R^T\AC R=F_0^T(P_0^T\AC P_0)F_0,
\eeq
the second term in $\AC _\eff$. As one can see, $P_0$ specifies the 
allowed form of the optical potential
while the {\bf form factor} $F_0$ specifies the coefficients in the 
optical potential, and thus introduces energy-dependent 
{\bf running coupling constants}.

The art in applying the reduction technique consists in finding 
embeddings $P$ that `dress' the subsystem of interest in a 
sufficiently accurate way,
a doorway operator $P_0$ embedding the relevant doorway states, 
and an operator $\widetilde \AC $ such that $U$ is small, 
and the limit \gzit{fo.2} exists and can be approximated efficiently.

Setting $P_0=0$ gives $R=0$, $\Delta=0$, and hence the trivial 
approximation 
\lbeq{fo.4a}
\AC _\eff\approx P^T\AC P.
\eeq
This simply amounts to discarding the interaction of the subsystem 
with the rest of the system. The choice $\Hz_0=\Hz$, $P_0=P$ is not 
better since then $P^TPF_0=P^TP_0F_0=P^TR=0$ by \gzit{pr.37}, and 
hence $F_0=0$, $R=0$. This is not surprising since we expect 
that the doorway operator $P_0$ should incorporate {\em additional} 
information about doorway states not yet represented in the subsystem 
but significantly interacting with it.

If $\Hz_0$ and $\Hz_\eff$ are finite-dimensional then \gzit{fo.1} 
defines finite-dimensional matrices, and the computation of the form 
factor amounts to solving the matrix equation 
\lbeq{fo.6}
   \left(\bary{ll}
          \AC _0 & P_1^T\\
          P_1 & 0
         \eary\right){F_0\choose F_1}={U\choose 0}
\eeq
with a complex symmetric (and for $\AC =\AC ^*$, $\bar P_0=P_0$ Hermitian)
coefficient matrix. 
\gzit{fo.6} can be solved efficiently by sparse matrix methods 
(cf. {\sc Duff} et al. \cite{ReiD,DufGR}) if suitable localized basis 
functions are used to construct $P$ and $P_0$.

In practice, it may be useful to employ in combination with 
discretization methods a complex absorbing potential in place of the 
$+i\eps$. In particular, if one proceeds as in 
{\sc Neumaier \& Mandelshtam} \cite{NeuM} one gets a quadratic 
eigenvalue problem that can handle all energies in a certain range 
simultaneously using harmonic inversion 
({\sc Mandelshtam \& Taylor} \cite{ManT}).

If $\Hz_\eff$ is finite-dimensional but $\Hz_0$ is a function space 
then $\AC _0$ is a differential or integral operator on $\Hz_0$. 
By solving suitable differential or integral equations we can find
the vector-valued functions
\[
B_0:=\lim_{\eps\downto 0}\AC _\eps^{-1}P_0^T\AC P,~~~
B_1:=\lim_{\eps\downto 0}\AC _\eps^{-1}P_0^TP.
\]
and the complex symmetric matrix 
\[
   G_1:=P_1B_1=\lim_{\eps\downto 0}(P_1\AC _\eps^{-1}P_1^T+i\eps).
\]
Noting that $\AC '=F_1$ if $\tilde \AC =0$, the formula \gzit{fo.3e} for the 
form factor becomes
\[
   F_0=B_0-B_1\AC ',~~~\mbox{where } \AC '=G_1^{-1}P_1B_0.
\]

\section{The quality of approximate state vectors}\label{s.qual}

In practice, it is usually impossible to find exact solutions of a 
Schr\"odinger equation $\AC \psi=0$. On the other hand, in real 
applications, $\AC $ is never precisely known either. Hence it 
makes sense to assess the quality of an approximate state vector 
$\psi$ by trying to modify $\AC $ a little to an operator $\widetilde \AC $ that 
satisfies $\widetilde \AC \psi=0$ exactly. 
If the modification $\widetilde \AC -\AC $ is within the accuracy to which $\AC $ 
is known, we are confident that $\psi$ is a good approximation to the 
true but unknown $\AC $.

This way of assessing the quality of an approximate solution of a 
problem is widely used in numerical analysis (see, e.g.,  
{\sc Wilkinson} \cite{Wil}) and is known under the name of 
{\bf backward error analysis}. Here we give a backward error analysis 
for the equation $\AC \psi=0$, and deduce guidelines for quality 
assessment of approximate state vectors. 

\begin{thm}
Let $\AC \in \Lin \Hz $ be symmetric. Then, for arbitrary $\psi\in \Hz^*$
with finite $\psi^T\psi\ne 0$, the modified operator
\lbeq{qu.2}
   \widehat \AC =\Big(1-{\psi\psi^T\over\psi^T\psi}\Big)\AC 
   \Big(1-{\psi\psi^T\over\psi^T\psi}\Big).
\eeq
satisfies $\widehat \AC \psi=0$ and
\lbeq{qu.1a}
 \tr(\widehat \AC -\AC )^2=\tau_\AC (\psi),
\eeq
where
\lbeq{qu.1aa}
 \tau_\AC (\psi):=2{(\AC \psi)^T(\AC \psi)\over \psi^T\psi}
 -\Big({\psi^T\AC \psi\over \psi^T\psi}\Big)^2.
\eeq
\end{thm}

\bepf
Since the formulas are invariant under scaling $\psi$ we may assume 
that $\psi$ is normalized to norm 1. Then 
\[
   \psi^T\psi=1,~~~\psi^T\AC \psi=:\lambda,~~~\psi^T\AC ^2\psi=:\mu,
\]
with real $\lambda$, $\mu$. The operator
\[
   \Delta:=\AC -\widehat \AC =\psi\psi^T\AC +\AC \psi\psi^T-\lambda\psi\psi^T
\]
satisfies $\Delta\psi=\AC \psi$, hence $\widehat \AC \psi=0$. Since
\[
\Delta \AC \psi=\lambda \AC \psi+(\mu-\lambda^2)\psi,
\]
we find
\[
  \Delta^2=\AC \psi\psi^T\AC +(\mu-\lambda^2)\psi\psi^T.
\]
$\Delta^2$ maps $\Hz$ to the two-dimensional space spanned by $\psi$ 
and $\AC \psi$, hence is trace class. Using the formula 
$\tr\phi\psi^T=\psi^T\phi$, we find
\[
\tr\Delta^2=\mu+(\mu-\lambda^2)=2\mu-\lambda^2=\tau_\AC (\psi), 
\]
giving \gzit{qu.1a}.
\epf

In the conservative case where $\im \AC =0$, and hence $\AC $ is Hermitian,
it is possible to show that the choice \gzit{qu.2} is best possible.
Note that a state vector $\psi\in\Hz^*$ with $\bar\psi=\psi$ and
finite $\psi^T\psi$ is now in the Hilbert space $\bar\Hz$.

\begin{thm}
Let $\AC \in \Lin \Hz $ be Hermitian, and suppose that
$\psi\in\bar\Hz\backslash\{0\}$ satisfies $\bar\psi=\psi$ 
and $\AC \psi\in\bar\Hz$.

(i) Any symmetric and Hermitian $\widetilde \AC \in \Lin \Hz $ with 
$\widetilde \AC \psi=0$ satisfies 
\lbeq{qu.1}
 \tr(\widetilde \AC -\AC )^2\geq\tau_\AC (\psi).
\eeq
Equality in \gzit{qu.1} is achieved precisely when 
$\widetilde \AC =\widehat \AC $. 

(ii) We always have 
\lbeq{qu.3}
0\le {(\AC \psi)^T\AC \psi\over\psi^T\psi}\leq\tau_\AC (\psi)\leq 2
     {(\AC \psi)^T\AC \psi\over\psi^T\psi}.
\eeq
In particular, $\tau_\AC (\psi)=0$ if and only if $\AC \psi=0$.
\end{thm}

\bepf
By the preceding theorem, $\widetilde \AC =\widehat \AC $ gives equality in 
\gzit{qu.1}, and is a good choice since $\widehat \AC \psi=0$. 
Hence suppose that $\widetilde \AC \neq\widehat \AC $. Without loss of 
generality, $(\widetilde \AC -\AC )^2$ is trace class (otherwise the trace 
is infinity and \gzit{qu.1} is trivially satisfied). Since 
$(\widetilde \AC -\widehat \AC )\psi=\widetilde \AC \psi-\widehat \AC \psi=0$ we 
have
\[
  \bary{ll}
   \tr(\AC -\widehat \AC )(\widetilde \AC -\widehat \AC ) 
   &=\tr\Delta(\widetilde \AC -\widehat \AC )\\
   &=\psi^T\AC (\widetilde \AC -\widehat \AC )\psi
    +\psi^T(\widetilde \AC -\widehat \AC )\AC \psi
    -\lambda\psi^T(\widetilde \AC -\widehat \AC )\psi=0.
  \eary
\]
Therefore 
\[
  \bary{lll}
   \tr(\widetilde \AC -\AC )^2-\tau_\AC (\psi) 
   &=&\tr(\widetilde \AC -\AC )^2-\tr(\widehat \AC -\AC )^2\\
   &=&\tr(\widetilde \AC +\widehat \AC -2\AC )(\widetilde \AC -\widehat \AC )\\
   &=&\tr(\widetilde \AC -\widehat \AC )^2
      -2\tr(\AC -\widehat \AC )(\widetilde \AC -\widehat \AC )\\
   &=&\tr(\widetilde \AC -\widehat \AC )^2>0
  \eary
\]
since $\widetilde \AC \neq\widehat \AC $. Therefore, \gzit{qu.1} holds for 
$\widetilde \AC \neq\widehat \AC $ with strict inequality. This proves (i). 
Since $\bar \psi=\psi$, the second term in \gzit{qu.1a} is nonnegative.
This gives the upper bound in \gzit{qu.3} and implies the final 
assertion. The lower bound follows from the Cauchy-Schwarz inequality 
$(\psi^T \AC \psi)^2\leq\psi^T\psi\cdot(\AC \psi)^T\AC \psi$.
\epf

\gzit{qu.2} is the orthogonal projection of $\AC $ to the orthogonal
complement of $\psi$, and $\sqrt{\tau_\AC (\psi)}$ measures, in a sense,
its deviation from $\AC $. Therefore, $\tau_\AC (\psi)$ (or its square root) 
serves as a useful measure for the quality of an approximate solution 
$\psi$ of the Schr\"odinger equation with Hermitian $\AC $.

In the nonhermitian case, it seems possible that
$\tau_\AC (\psi)=0$ even if $\AC \psi\ne 0$. (A finite-dimensional example
is $\AC ={ \alpha~~i \choose i~~0}$, $\psi={1 \choose 0}$ which has
$\AC ^2=0$ and $\tau_\AC (\psi)=0$ for $\alpha=\sqrt{2}$, and for 
$\alpha=1$ even $\tau_\AC (\psi)<0$.) Thus, unless $\AC ^*=\AC $, the measure 
$\tau_\AC (\psi)$ might be sometimes too optimistic.
However, in practice $\AC $ is nearly Hermitian and the use of
$\tau_\AC (\psi)$ should cause no problems.

\section{Multichannel scattering}\label{s.multi}

We now apply the above to the multichannel approach discussed 
below. A projection approach to multichannel scattering leading to
effective Hamiltonians is discussed, e.g., in {\sc Newton} 
\cite[Section 16.6]{New}, but the equations derived there  
appear not to be suitable to numerical approximation. A more useful
formulation is given by the present equations from Theorem \ref{t.form},
with $P$ and $P_0$ as given below.

An {\bf arrangement} is a partition $A$ of the system of particles into 
clusters $i\in A$, with correct assignment of distinguishability.
At a fixed energy $E$, those arrangements are relevant that contain
{\bf channels} defined by cluster bound states with energies $E_i$ 
such that 
\[
\sum_{i\in A} E_i \le E+\Delta E
\]
where $\Delta E$ is zero or a small quantity. These open or nearly 
open channels are assumed to correspond approximately to states 
in a $n_A$-dimensional space 
\[
\Hz_{A0}\subseteq \bigotimes_{i\in A} \Hz_i
\] 
with basis functions
\lbeq{e.basis}
\phi_{Ak}(x)=\prod_{i\in A} \phi_{il_{ik}}(x_i),
\eeq
where $x_i$ is the vector of coordinates of particles in cluster $i$,
and the $\phi_{il}(x_i)$ are translation invariant basis functions 
from the cluster Hilbert space $\Hz_i$, used in all possible combinations in \gzit{e.basis}. The motion of the clusters 
is described by a space $\Hz_A$ of functions of a system of relative 
coordinates $r_A$ between the cluster centers.

Consider, for example, a 3-particle reaction 
$XY+Z\rightleftharpoons X+YZ$.
Then $\Hz_{XY+Z,0}$ consists of products of approximate bound states 
$\psi_{XY}$ and the ground state $\psi_Z$; $\Hz_{X+YZ,0}$ consists of 
products of the ground state $\psi_X$ and approximate bound states 
$\psi_{YZ}$, and $\Hz_{XYZ,0}$ consists of sufficiently many states 
localized in the transition region to resolve the resonances of 
interest. Usually, one would keep the arrangements $(XY,Z)$ and 
$(X,YZ)$ in the reduced description, and use the states belonging to 
the arrangement $(XYZ)$ as doorway states for the transition regime.

Let ${\cal C}$ be the set of arrangements considered relevant for the 
reduced description. The reduced 
{\bf multichannel state space} is then the dual of the space 
\[
\Hz_\eff=\bigoplus_{A\in {\cal C}_{\eff}}\Hz_A^{n_A}
\]
or the properly symmetrized subspace in case of indistinguishable 
clusters. The inner product in $\Hz_\eff$ is given by 
\[
\phi_\eff^T\psi_\eff=
\sum_{A\in{{\cal C}_{\eff}}}\int dr_A \phi_A(r_A)^T\psi_A(r_A).
\]
The embedding map $P:\Hz_\eff^*\to\Hz^*$ is given by
\[
   P\psi_\eff=\sum_{A\in{{\cal C}_{\eff}}}P_A\psi_A,
\]
where the $k$th component of $P_A:(\Hz_A^{n_A})^*\to\Hz^*$ maps a 
function of $r_A$ to the $k$th basis vector of $\Hz_A$ modified to 
have the corresponding dependence on the center of mass $r_A(x)$
of the cluster coordinates $x$,
\[
(P_A\psi_A)_k(x)=\psi_A(r_A(x))\phi_{Ak}(x).
\]
The transpose 
$P^T:\Hz^*\to\Hz_\eff^*$ is given by
\[
   (P^T\psi)_A=P_A^T\psi~~~\mbox{for all}~A\in{{\cal C}_{\eff}}.
\]
By construction, $P^T\AC P$ is a direct sum of contributions of the form
\[
P_A^T\AC P_A = \AC _A-H_A,
\]
where $H_A$ is the free Hamiltonian for the motion of the cluster 
centers and $\AC _A$ is a symmetric $n_A\times n_A$-matrix. In a basis of
cluster eigenstates with energies $E_i$, $\AC _A$ is the diagonal matrix 
formed by the energy differences $E-E_i$. 

The construction of $(\Hz_0,P_0)$ is completely
analogous, using a larger set of arrangements and/or channels that 
contain the doorway states.
  
Thus the calculations have the same 
complexity as those for the coupled reaction channel equations 
(or resonating group method), as described, e.g., by
{\sc Wildermuth \& Tang} \cite{WilT}). However, the present scheme is 
more flexible in that it can incorporate information from doorway 
states. In principle, by increasing the size of the doorway
state space, it is capable of arbitrarily accurate approximations to
the full dynamics, and shares this feature with the two Hilbert space 
method of {\sc Chandler \& Gibson} \cite{ChaG,ChaG2} and with a
technique by {\sc Goldflam \& Kowalski} \cite{GolK}.

To solve the reduced Schr\"odinger equation (and, if a similar 
construction is used for the doorway operator, the equations 
for the form factor), the whole arsenal of methods developed 
in the applications is available. Binary arrangements can be 
handled by Lippmann-Schwinger equations (see, e.g., 
{\sc Wildermuth \& Tang} \cite{WilT}, 
{\sc Adhikari \& Kowalski} \cite[Chapter 3]{AdhK}), 
and 3-cluster arrangements by 
the {\sc Faddeev} \cite{FadM} connected kernel approach 
(see, e.g., {\sc Gl\"ockle} \cite[Chapter 3]{Glo}, 
{\sc Kukulin} et al. \cite{KukKH},
{\sc Adhikari \& Kowalski} \cite[Chapter 7]{AdhK}).


\end {document}